\documentclass[preprint,12pt]{elsarticle}

\usepackage{amssymb}
\usepackage{graphicx}
\usepackage{amsmath}
\usepackage{amsfonts}

\begin{document}

\begin{frontmatter}



\title{Variational integrators for perturbed non-canonical Hamiltonian systems}


\author[pppl]{J. W. Burby}
\author[pppl]{C. L. Ellison}
\author[pppl,ustc]{H. Qin}

\address[pppl]{Princeton Plasma Physics Laboratory, Princeton, New Jersey 08543, USA}
\address[ustc]{Dept. of Modern Physics, University of Science and Technology of China, Hefei, Anhui 230026, China}

\begin{abstract}
Finite-dimensional non-canonical Hamiltonian systems arise naturally from Hamilton's principle in phase space. We present a method for deriving variational integrators that can be applied to perturbed non-canonical Hamiltonian systems on manifolds based on discretizing this phase-space variational principle. Relative to the perturbation parameter $\epsilon$, this type of integrator can take $O(1)$ time steps with arbitrary accuracy in $\epsilon$ by leveraging the unperturbed dynamics. Moreover, these integrators are coordinate independent in the sense that their time-advance rules transform correctly when passing from one phase space coordinate system to another. 
\end{abstract}

\begin{keyword}
variational integrators \sep geometric mechanics \sep perturbation theory

\end{keyword}

\end{frontmatter}



\section{Introduction} 
The most famous variational principle in classical mechanics is \emph{Hamilton's principle} of stationary action \cite{Landau_1976}. According to Hamilton's principle, a system's path in configuration space, $q(t)\in Q$, will be a critical point of 
\begin{align}
\mathcal{A}(q)=\int_{t_1}^{t_2}\!L(q(t),\dot{q}(t),t)\,dt\,
\end{align}  
regarded as a functional of paths in configuration space with fixed endpoints.
Here, $L$ is the Lagrangian function associated with the mechanical system in question. The closely-related \emph{Hamilton's principle in phase space} \cite{Arnold_1989} generalizes Hamilton's principle to arbitrary exact symplectic manifolds. Specifically, if the symplectic manifold $M$ with symplectic form \cite{Abraham_2008} $-\mathbf{d}\vartheta$ is the phase space of a mechanical system, then the phase space variational principle asserts that a system's path in phase space, $z(t)\in M$, will be a critical point of  
\begin{align}\label{phase_space}
S(z)=\int_{t_1}^{t_2}\!\vartheta_{z(t)}(\dot{z}(t))-H(z(t),t)\,dt\,
\end{align}
regarded as a functional of paths in phase space with fixed endpoints.
Here, $H$ is the system's Hamiltonian. This pair of variational principles serves as the variational workhorse of modern treatments of mechanics. Hamilton's principle is often well-suited to the formulation of a problem, as evidenced by its applications in continuum mechanics \cite{Holm_1998}, whereas the phase space principle is naturally adapted to perturbation theory, especially in guiding center and gyrokinetic theory \cite{Brizard_2007, Cary_2009}. 

In this article, we will investigate variational discretizations \cite{Marsden_2001} of \emph{perturbed} non-canonical Hamiltonian systems on manifolds that obey Hamilton's principle in phase space. These perturbed systems will be specified by a phase space manifold $M$; a symplectic form on $M$ of the form $-\mathbf{d}\vartheta$; and a time-dependent Hamiltonian function $\mathcal{H}_t=H_{t}+\epsilon h_t$, where $H_{t}$ represents an unperturbed system and $\epsilon$ is a small parameter. Trajectories of this type of system are then given as extremals of the action $S$ given in Eq.\,(\ref{phase_space}). A wide variety of mechanical systems fit this mold, including all perturbed canonical Hamiltonian systems. Notably, there are perturbed Hamiltonian systems for which Hamilton's principle in phase space is the only known variational formulation \cite{Cary_2009,Aref_2007}. The variational discretizations we will be concerned with are those that exploit the small value of $\epsilon$ to enhance the accuracy of the discrete Euler-Lagrange equations; see Refs. \cite{Qin_2008,Qin_2009,Rowley_2002,Li_2011} for generic variational discretizations of Hamilton's principle in phase space and Refs. \cite{Lall_2006, Leok_2011} for specialized methods that can be applied to Hamilton's principle in phase space while working in canonical coordinates.  

Discretizations of Hamilton's principle for perturbed systems have been developed already in Ref.\,\cite{Farr_2009}. The discrete Euler-Lagrange equations associated with these discretizations are capable of recovering previously-discovered symplectic integrators \cite{Mclachlan_1995, Chambers_2000, Laskar_2001} that exploit the small value of $\epsilon$ to effectively enhance their order of accuracy. In particular, these integrators are capable of achieving up to local $O(\epsilon^2)$ accuracy (how to achieve greater accuracy in $\epsilon$ with this type of integrator is not discussed in Ref.\,\cite{Farr_2009}). On the other hand, it seems a method for finding discretizations of Hamilton's principle in phase space that exploits the small value of $\epsilon$ has never been discussed. One might hope that the technique used in Ref.\,\cite{Farr_2009} could be easily extended to treat the phase space variational principle, but this is not the case. The derivation of the discretizations in Ref.\,\cite{Farr_2009} depends on the presence of an exact discrete Lagrangian for its success. Because the usual notion of exact discrete Lagrangian\,\cite{Marsden_2001} only applies to Hamilton's principle\footnote{The exact discrete Lagrangian associated with Hamilton's principle is a function of pairs of points in configuration space. The phase-space action cannot be regarded as a function of pairs of points in phase space  in the same way because generally there is not a solution to Hamilton's equations that connects a pair of points in phase space. }, this method cannot be transcribed to produce a similar method for discretizing Hamilton's principle in phase space. Thus, there is currently not a method for developing variational integrators for perturbed guiding center motion, or any other perturbed system whose only known variational formulation is in the form of Hamilton's principle in phase space. 

The purpose of this article is to formulate the first discretizations of Hamilton's principle in phase space that are adapted to perturbed problems. These discretizations are contained in our Eqs.\,(\ref{l_infinity}),\,(\ref{l_0}),\,(\ref{l_1}), and\,(\ref{l_2}). Amongst them are discretizations whose time-advance rules are accurate to any desired order in $\epsilon$ while allowing for $O(1)$ time steps. Each discretization can be applied to any perturbed non-canonical Hamiltonian system whose continuous-time trajectories are extremals of the action $S$ in Eq.\,(\ref{phase_space}). Our method for deriving these discretizations is based on constructing an exact discrete action for Hamilton's principle in phase space, a task that is fundamentally different from constructing an exact discrete action for Hamilton's principle. The method is completely coordinate-independent, and therefore leads to time-advance rules that transform correctly when passing from one coordinate chart on the phase space $M$ to another. This coordinate-independence is achieved by introducing an arbitrary affine connection on $M$. 
We demonstrate that by initializing these two-step integration algorithms using the smooth modified system studied in backward error analysis \cite{Hairer_1999, Hairer_2006, Ellison_2013}, the resulting discrete-time trajectories satisfy one-step algorithms that preserve symplectic forms on $M$, which is a sharper result on symplecticity than that provided by the theory developed in Ref. \cite{Marsden_2001} or Ref. \cite{Rowley_2002}. 

The presentation will be organized as follows. We specify the scope of our work and precisely define the notion of a discretization of Hamilton's principle in phase space in Section \ref{sec:problem_statement}. We derive an \emph{exact} discretization of Hamilton's principle in phase space suitable to perturbed Hamiltonian systems in Section \ref{sec:derivation_of_l_infinity}. Using this exact discretization, we develop approximate discretizations of Hamilton's principle in phase space that can be applied to practical problems in Section \ref{sec:truncations_of_l_infinity}. In section \ref{sec:symplecticity}, we discuss the symplecticity of the integration algorithms provided by our discretizations. Section \ref{sec:examples} contains two examples. We conclude with a discussion in Section \ref{sec:discussion}.

\section{\label{sec:problem_statement}Problem statement}
Let $M$ be a symplectic manifold with symplectic form $-\mathbf{d}\vartheta$. Let $\mathcal{H}_t=H_t+\epsilon h_t$ be a time-dependent real-valued function on $M$, where $\epsilon$ is a small parameter. The time-dependent vector field $X_{\mathcal{H}_t}$ defined by the formula
\begin{align}
\mathbf{i}_{X_{\mathcal{H}_t}}\mathbf{d}\vartheta=-\mathbf{d}\mathcal{H}_t
\end{align}
is known as the time-dependent Hamiltonian vector field with Hamiltonian $\mathcal{H}_t$ \cite{Abraham_2008}. Fix $t_1,t_2\in\mathbb{R}$ with $t_1<t_2$. If $\gamma:[t_1,t_2]\rightarrow M$ is an integral curve of $X_{\mathcal{H}_t}$, i.e.
\begin{align}
\gamma^\prime(t)=X_{\mathcal{H}_t}(\gamma(t)),
\end{align}
then $\gamma$ is a critical point of the functional $S_{(\gamma(t_1),\gamma(t_2))}:\mathcal{P}_{(\gamma(t_1),\gamma(t_2))}\rightarrow\mathbb{R}$, where
\begin{align}
\mathcal{P}_{(z_1,z_2)}=\{c:[t_1,t_2]\rightarrow M\,|\,c(t_1)=z_1,\,c(t_2)=z_2\}
\end{align}
and
\begin{align}
S_{(\gamma(t_1),\gamma(t_2))}(c)=\int_{t_1}^{t_2}\vartheta_{c(t)}(c^\prime(t))-\mathcal{H}_t(c(t))\,dt .
\end{align}
Conversely, if $c\in\mathcal{P}_{(z_1,z_2)}$ is a critical point of $S_{(z_1,z_2)}$, then $c$ must be an integral curve of the Hamiltonian vector field $X_{\mathcal{H}_t}$. The latter pair of facts is known as Hamilton's principle in phase space \cite{Arnold_1989}. Note that for many choices of $(z_1,z_2)$, $S_{(z_1,z_2)}$ will not have \emph{any} critical points; a necessary condition for the existence of a critical point is that $z_1$ and $z_2$ can be connected by an integral curve of $X_{\mathcal{H}_t}$.

Let $\tau\in\mathbb{R}$ be a positive $O(1)$ constant. Assume $t_1=N_1\tau$ and $t_2=N_2\tau$, where $N_1,N_2\in\mathbb{Z}$. Between the continuous-time path space, $\mathcal{P}_{(z_1,z_2)}$, and the discrete-time path space with increment $\tau$,
\begin{align}
\mathbb{P}_{(z_1,z_2)}=\{\mathbf{c}:[t_1,t_2]\cap(\tau\mathbb{Z})\rightarrow M\,|\,\mathbf{c}(t_1)=z_1,\,\mathbf{c}(t_2)=z_2\},
\end{align} 
there is a natural projection map $\pi_{(z_1,z_2)}:\mathcal{P}_{(z_1,z_2)}\rightarrow\mathbb{P}_{(z_1,z_2)}$, where for each integer $N_1\leq k\leq N_2$
\begin{align}
\pi_{(z_1,z_2)}(c)(k\tau)=c(k\tau).
\end{align}
Our goal is to identify a functional $\mathbb{S}^\infty_{(z_1,z_2)}:\mathbb{P}_{(z_1,z_2)}\rightarrow\mathbb{R}$, which we will refer to as the exact discrete action, with three properties. 
\\ \\
(D1) If $\gamma:[t_1,t_2]\rightarrow M$ is an integral curve of the Hamiltonian vector field $X_{\mathcal{H}_t}$ with $\gamma(t_1)=z_1$ and $\gamma(t_2)=z_2$, then $\gamma$ must be a critical point of the functional $\mathbb{S}^\infty_{(z_1,z_2)}\circ\pi_{(z_1,z_2)}$. 
\\ \\
(D2) $\mathbb{S}^\infty_{(z_1,z_2)}$ must be of the form
\begin{align}
\mathbb{S}^\infty_{(z_1,z_2)}(\mathbf{c})=\sum_{k=N_1}^{N_2-1}L_\infty(\mathbf{c}_k,\mathbf{c}_{k+1},\tau k),
\end{align}
where $\mathbf{c}_k=\mathbf{c}(\tau k)$ and $L_\infty:M\times M\times\mathbb{R}\rightarrow\mathbb{R}$.
\\ \\
(D3) If the exact discrete action is replaced with its $l$'th order Maclaurin polynomial in $\epsilon$,
\begin{align}
\mathbb{S}^{l}_{(z_1,z_2)}(\mathbf{c})=\sum_{k=N_1}^{N_2-1}L_l(\mathbf{c}_k,\mathbf{c}_{k+1},\tau k),
\end{align}
then the discrete Euler-Lagrange equations \cite{Marsden_2001} associated with $\mathbb{S}^l_{(z_1,z_2)}$ should function as a numerical integration algorithm with local $O(\epsilon^{l+1})$ accuracy. 

Remarks on such a functional are in order. Note that the time step of the integration algorithm associated with $\mathbb{S}^l_{(z_1,z_2)}$ is $\tau$, which is not assumed to be small. The idea at work here is that integral curves of the unperturbed vector field $X_{H_t}$ can be used to approximate integral curves of $X_{\mathcal{H}_t}$ with $O(\epsilon)$ accuracy on $O(1)$ time intervals. Thus, when $H_t$ describes an integrable Hamiltonian system, the practical limit on the size of $\tau$ for a fixed value of $\epsilon$ should be expected to be $\tau\ll\tau_b$, where $\tau_b$ is the perturbation's characteristic bounce time. Also note that because the discrete Euler-Lagrange equations associated with $\mathbb{S}^l_{(z_1,z_2)}$ are given by
\begin{align}\label{introDEL}
\mathbf{d}^{(2)}L_l(\mathbf{c}_{k-1},\mathbf{c}_k,\tau (k-1))+\mathbf{d}^{(1)}L_l(\mathbf{c}_k,\mathbf{c}_{k+1},\tau k)=0,
\end{align}
they provide a \emph{two-step} \cite{Dahlquist_1956, Hairer_1999, Hairer_2006} integration algorithm for a first-order dynamical system on $M$. Thus, the standard theory behind variational integrators \cite{Marsden_2001} implies that this algorithm preserves a symplectic structure on $M\times M$. However, the flow of the Hamiltonian vector field $X_{\mathcal{H}_t}$  preserves a symplectic form on $M$. In Section \ref{sec:symplecticity} we reconcile these qualitatively different notions of structure preservation using a minor modification of the ``smooth modified system" concept developed in \cite{Hairer_1999}. Finally, note that by setting $\epsilon=0$, we move into the setting of generic Hamiltonian systems on $M$, i.e. those without an \emph{a priori} perturbative structure. Thus, when $\epsilon=0$ and $\tau$ is chosen to be small, $\mathbb{S}^\infty_{(z_1,z_2)}$ can be expanded in powers of $\tau$ to yield arbitrarily accurate variational integrators for generic Hamiltonian systems.

\section{\label{sec:derivation_of_l_infinity}Derivation of an exact discrete action for Hamilton's principle in phase space}
In order to derive a functional $\mathbb{S}^\infty_{(z_1,z_2)}$ that satisfies properties (D1)--(D3), we will manipulate the functional $S_{(z_1,z_2)}$ into the form $\mathbb{S}^\infty_{(z_1,z_2)}\circ\pi_{(z_1,z_2)}$ while making use of the following heuristic approximation principle.
\\ \\
\noindent{\bf The path space approximation principle:} Modifications to the functional $S_{(z_1,z_2)}$ can be made as long as they do not change the first variation of $S_{(z_1,z_2)}$ at integral curves of $X_{\mathcal{H}_t}$.
\\ \\
\noindent The intuitive justification of this principle is that we are mainly interested in critical points of the functional $S_{(z_1,z_2)}$ and a critical point of $S_{(z_1,z_2)}$ will also be a critical point of $S^\prime_{(z_1,z_2)}$ provided these two functionals agree (modulo a constant) in a neighborhood of the critical point. 

As a convenient first step, we will pass into the ``interaction picture". Passing into the interaction picture amounts to transforming the path space in such a way that integral curves of the unperturbed vector field $X_{H_t}$ become trivially constant curves. Let $F_{t,s}:M\rightarrow M$ be the time-dependent flow map of the unperturbed Hamiltonian vector field $X_{H_t}$, i.e. the two-parameter family of mappings characterized by the relations
\begin{align}
F_{t,t}(z)&=z\\
\frac{\mathrm{d}}{\mathrm{d}t}F_{t,s}(z)&=X_{H_t}(F_{t,s}(z)).
\end{align} 
Then the interaction picture transformation from the old path space to the new path space $\mathcal{I}:\mathcal{P}_{(z_1,z_2)}\rightarrow \mathcal{P}_{(\bar{z}_1,\bar{z}_2)}$, with $(\bar{z}_1,\bar{z}_2)=(z_1,F_{t_1,t_2}(z_2))$, is given by 
\begin{align}
\mathcal{I}(c)(t)=F_{t_1,t}(c(t)).
\end{align}
After performing this change-of-path, the functional $S_{(z_1,z_2)}$ is transformed into $\bar{S}_{(\bar{z}_1,\bar{z}_2)}=\mathcal{I}_*S_{(z_1,z_2)}$, which is the pushforward of $S_{(z_1,z_2)}$ along the mapping $\mathcal{I}$. Given $\bar{c}\in\mathcal{P}_{(\bar{z}_1,\bar{z}_2)}$, $\bar{S}_{(\bar{z}_1,\bar{z}_2)}(\bar{c})$ is readily found to be
\begin{align}
\bar{S}_{(\bar{z}_1,\bar{z}_2)}(\bar{c})=\int_{t_1}^{t_2}\vartheta_{\bar{c}(t)}(\bar{c}^\prime(t))-\epsilon (F^*_{t,t_1}h_t)(\bar{c}(t))\,dt+\text{const}.
\end{align}
The constant term does not affect the location of critical points, and so we omit it from this point forward. Also note that, by Hamilton's principle in phase space, $\bar{c}$ is a critical point of $\bar{S}_{(\bar{z}_1,\bar{z}_2)}$ if and only if $\bar{c}$ is an integral curve of the time-dependent Hamiltonian vector field $X_{\epsilon K_t}$, where
\begin{align}
K_t=F^*_{t,t_1}h_t.
\end{align}

Next, we decompose the time integral in $\bar{S}_{(\bar{z}_1,\bar{z}_2)}$ as
\begin{align}\label{decomp}
\bar{S}_{(\bar{z}_1,\bar{z}_2)}(\bar{c})=\sum_{k=N_1}^{N_2-1}\bar{S}^k(\bar{c}) ,
\end{align}
where
\begin{align}
\bar{S}^k(\bar{c})=\int_{\tau k}^{\tau(k+1)}\vartheta_{\bar{c}(t)}(\bar{c}^\prime(t))-\epsilon K_t(\bar{c}(t))\,dt,
\end{align}
and examine $\bar{S}^k(\bar{c})$ for each $k$. The goal of this analysis is to devise an approximation for $\bar{S}^k(\bar{c})$ that depends on $\bar{c}$ only through $\bar{c}_k$ and $\bar{c}_{k+1}$. Let $G_{t,s}:M\rightarrow M$ be the time-dependent flow map of the Hamiltonian vector field $X_{\epsilon K_t}$. As is readily verified, the identity
\begin{align}
\int_{\tau k}^{\tau(k+1)}\vartheta_{\bar{c}(t)}(\bar{c}^\prime(t))-\epsilon K_t(\bar{c}(t))\,dt=\int\limits_{\bar{\bar{c}}^k}\vartheta&+\bigg(\epsilon\int\limits_{\tau(k+1/2)}^{\tau(k+1)} G^*_{s,\tau(k+1)}l_s\,ds\bigg)(\bar{c}_{k+1})\nonumber\\
&+\bigg(\epsilon\int\limits_{\tau k}^{\tau(k+1/2)}G^*_{s,\tau k} l_s\,ds\bigg)(\bar{c}_k)
\end{align}
holds, where
\begin{align}
\bar{\bar{c}}^k(t)=G_{\tau(k+1/2),t}(\bar{c}(t)),~\tau k\leq t\leq \tau(k+1)
\end{align}
and
\begin{align}
l_s=\vartheta(X_{K_s})-K_s.
\end{align} 
Thus, $\bar{S}^k(\bar{c})$ could be determined using only the values of $\bar{c}$ at $\tau k$ and $\tau(k+1)$ were it not for the term $\int_{\bar{\bar{c}}^k}\vartheta$. 

Observe that when $\bar{c}$ is a critical point of $\bar{S}_{(\bar{z}_1,\bar{z}_2)}$, the curve $\bar{\bar{c}}^k$ is constant and the integral $\int_{\bar{\bar{c}}^k}\vartheta$ vanishes. When $\bar{c}$ is infinitesimally close to a critical point, $\int_{\bar{\bar{c}}^k}\vartheta$ is given by the first variation of the functional $\mathcal{F}_1:\mathcal{P}_{(\bar{z}_1,\bar{z}_2)}\rightarrow\mathbb{R}$, where
\begin{align}
\mathcal{F}_1(\bar{c})=\int\limits_{\bar{\bar{c}}^k}\vartheta.
\end{align} 
To calculate the variation of $\mathcal{F}_1$, we first compute the variation of $\mathcal{F}_o:\mathcal{P}_{(\bar{z}_1,\bar{z}_2)}\rightarrow\mathbb{R}$ at a trivial curve $\bar{c}(t)=\text{const}$, where
\begin{align}
\mathcal{F}_o(\bar{c})=\int\limits_{\bar{c}^k}\vartheta,
\end{align}
and $\bar{c}^k=\bar{c}|[\tau k,\tau(k+1)]$. 
The result is readily found to be
\begin{align}
\mathbf{d}\mathcal{F}_{o\,\bar{c}}(\delta\bar{c})=\vartheta_{\bar{c}_{k+1}}(\delta\bar{c}_{k+1})-\vartheta_{\bar{c}_k}(\delta\bar{c}_k).
\end{align}
The chain rule then implies that the first variation of $\mathcal{F}_1$ at a critical point of $\bar{S}_{(\bar{z}_1,\bar{z}_2)}$ is given by
\begin{align}\label{first_variation}
\mathbf{d}\mathcal{F}_{1\,\bar{c}}(\delta\bar{c})=&\vartheta_{G_{\tau(k+1/2),\tau(k+1)}(\bar{c}_{k+1})}(TG_{\tau(k+1/2),\tau(k+1)}(\delta\bar{c}_{k+1}))\nonumber\\
-&\vartheta_{G_{\tau(k+1/2),\tau k}(\bar{c}_k)}(TG_{\tau(k+1/2),\tau k}(\delta\bar{c}_k)).
\end{align}
Here, $T$ denotes the tangent functor as defined in Ref.\,\cite{Abraham_2008}. Notably, the first variation of $\mathcal{F}_1$ at a critical point of $\bar{S}_{(\bar{z}_1,\bar{z}_2)}$ is completely determined by $\delta\bar{c}$ and $\bar{c}$ evaluated at $\tau k$ and $\tau (k+1)$.

Guided by these observations and the path space approximation principle, we will now replace the term $\int_{\bar{\bar{c}}^k}\vartheta$ in $\bar{S}^k(\bar{c})$ with an approximation that can be computed using only the values of $\bar{c}$ at $\tau k$ and $\tau(k+1)$. We will choose this approximation so that it agrees with $\int_{\bar{\bar{c}}^k}\vartheta$ when $\bar{c}$ is a critical point and when $\bar{c}$ is infinitesimally close to a critical point. Let $\nabla$ be an arbitrary affine connection on $M$. It is well-known \cite{Postnikov_2001} that $M$ admits an open cover $\{U_i\}_{i\in I}$ with two properties: (i) if $z_1,z_2\in U_i$, then there is a unique geodesic segment contained in $U_i$ with endpoints $z_1$ and $z_2$, (ii) If $z_1,z_2\in U_i$ and $z_1,z_2\in U_j$, then the geodesic segment joining $z_1,z_2$ in $U_i$ is equal to the geodesic segment joining $z_1,z_2$ in $U_j$. Thus, on the open neighborhood of the diagonal in $M\times M$, $\mathcal{O}=\bigcup\limits_{i\in I}\,U_i\times U_i$,
we can define a real-valued function
\begin{align}\label{f}
f(z_1,z_2)=\int\limits_{I(z_2,z_1)}\vartheta,
\end{align}
where $I(z_2,z_1)$ is the unique directed geodesic segment from $z_1$ to $z_2$ contained in some $U_i$. If $\mathcal{O}$ cannot be taken to be all of $M\times M$, assume that $f$ has been smoothly extended to all of $M\times M$. In terms of this possibly-extended $f$ our approximation for $\int_{\bar{\bar{c}}^k}\vartheta$ is
\begin{align}\label{fundamental_approximation}
\int\limits_{\bar{\bar{c}}^k}\vartheta\approx &f(F_{\tau k,t_1}(\bar{\bar{c}}_k),F_{\tau k,t_1}(\bar{\bar{c}}_{k+1}))\nonumber\\
&-\bigg(\int\limits_{t_1}^{\tau k}F^*_{s,t_1}\mathcal{L}_s\,ds\bigg)(\bar{\bar{c}}_{k+1})\nonumber\\
&+\bigg( \int\limits_{t_1}^{\tau k}F^*_{s,t_1}\mathcal{L}_s\,ds \bigg)(\bar{\bar{c}}_k),
\end{align}
where
\begin{align}
\mathcal{L}_s=\vartheta(X_{H_s})-H_s.
\end{align}
It is readily verified that this approximation is exact when $\bar{c}$ is either a critical point of $\bar{S}_{(\bar{z}_1,\bar{z}_2)}$ or infinitesimally close to such a critical point. When $\bar{c}$ is a critical point, both sides of Eq.\,(\ref{fundamental_approximation}) obviously vanish. Likewise, regarding each side of Eq.\,(\ref{fundamental_approximation}) as a functional of $\bar{c}$, the two sides' first variations at a critical point agree. The latter assertion is easy to check using Eq.\,(\ref{first_variation}) and the fact that when $\bar{c}$ is sufficiently close to a critical point, the right hand-side of Eq.\,(\ref{fundamental_approximation}) is given by
\begin{align}
\int\limits_{I(F_{\tau k,t_1}(\bar{\bar{c}}_{k+1}),F_{\tau k,t_1}(\bar{\bar{c}}_k))}F^*_{t_1,\tau k}\vartheta.
\end{align}

With this approximation in place, we can now easily obtain an expression in the interaction picture for a functional $\bar{\mathbb{S}}^\infty_{(\bar{z}_1,\bar{z}_2)}$ that satisfies (D1) and (D2). To see this, note that we now have $\bar{S}^k(\bar{c})\approx \bar{L}_\infty(\bar{c}_k,\bar{c}_{k+1},\tau k)$, where
\begin{align}
\bar{L}_\infty(\bar{c}_k,\bar{c}_{k+1},t)= &f(F_{t,t_1}(\bar{\bar{c}}_k),F_{t,t_1}(\bar{\bar{c}}_{k+1}))\nonumber\\
&-\bigg(\int\limits_{t_1}^{t}F^*_{s,t_1}\mathcal{L}_s\,ds\bigg)(\bar{\bar{c}}_{k+1})\nonumber\\
&+\bigg( \int\limits_{t_1}^{t}F^*_{s,t_1}\mathcal{L}_s\,ds \bigg)(\bar{\bar{c}}_k)\nonumber\\
&+\bigg(\epsilon\int\limits_{t+\tau/2}^{t+\tau} G^*_{s,t+\tau}l_s\,ds\bigg)(\bar{c}_{k+1})\nonumber\\
&+\bigg(\epsilon\int\limits_{t}^{t+\tau/2}G^*_{s,t} l_s\,ds\bigg)(\bar{c}_k).
\end{align}
Thus, Eq.\,(\ref{decomp}) implies 
\begin{align}
\bar{S}_{(\bar{z}_1,\bar{z}_2)}(\bar{c})\approx&\sum_{k=N_1}^{N_2-1}\bar{L}_\infty(\bar{c}_k,\bar{c}_{k+1},\tau k)=\bar{\mathbb{S}}^\infty_{(\bar{z}_1,\bar{z}_2)}\circ\pi_{(\bar{z}_1,\bar{z}_2)}(\bar{c}),
\end{align}
where $\bar{\mathbb{S}}^\infty_{(\bar{z}_1,\bar{z}_2)}:\mathbb{P}_{\bar{z}_1,\bar{z}_2}\rightarrow\mathbb{R}$ is given by
\begin{align}
\bar{\mathbb{S}}^\infty_{(\bar{z}_1,\bar{z}_2)}(\bar{\mathbf{c}})=\sum_{k=N_1}^{N_2-1}\bar{L}_\infty(\bar{\mathbf{c}}_k,\bar{\mathbf{c}}_{k+1},\tau k).
\end{align}
This says that $\bar{\mathbb{S}}^{\infty}_{(\bar{z}_1,\bar{z}_2)}$ satisfies (D2). Also note that, by construction, we have the following equalities when $\bar{c}$ is a critical point of $\bar{S}_{(\bar{z}_1,\bar{z}_2)}$.
\begin{align}
\bar{S}_{(\bar{z}_1,\bar{z}_2)}(\bar{c})&=\bar{\mathbb{S}}^{\infty}_{(\bar{z}_1,\bar{z}_2)}\circ\pi_{(\bar{z}_1,\bar{z}_2)}(\bar{c})\\
\mathbf{d}\bar{S}_{(\bar{z}_1,\bar{z}_2)\,\bar{c}}&=\mathbf{d}\left(\bar{\mathbb{S}}^{\infty}_{(\bar{z}_1,\bar{z}_2)}\circ\pi_{(\bar{z}_1,\bar{z}_2)}\right)_{\bar{c}}.
\end{align}
Therefore, any critical point of $\bar{S}_{(\bar{z}_1,\bar{z}_2)}$ is also a critical point of $\bar{\mathbb{S}}^{\infty}_{(\bar{z}_1,\bar{z}_2)}\circ\pi_{(\bar{z}_1,\bar{z}_2)}$, which says that $\bar{\mathbb{S}}^{\infty}_{(\bar{z}_1,\bar{z}_2)}$ satisfies (D1). 

In fact $\bar{\mathbb{S}}^{\infty}_{(\bar{z}_1,\bar{z}_2)}$ satisfies (D3) in addition to (D1) and (D2). The proof of this statement is not substantially different than the proof of Theorem 2.3.1 in Ref.\,\cite{Marsden_2001}. We have therefore succeeded in identifying an exact discrete action $\bar{\mathbb{S}}^\infty_{(\bar{z}_1,\bar{z}_2)}$. However, we currently have this action expressed in the interaction picture. To remedy this, we will conclude this section by passing out of the interaction picture. 

Passing out of the interaction picture consists of transforming the new path space $\mathcal{P}_{(\bar{z}_1,\bar{z}_2)}$ back into the old path space $\mathcal{P}_{(z_1,z_2)}$ by applying the mapping $\mathcal{I}^{-1}$. Upon performing this change-of-path, the functional $\bar{\mathbb{S}}^{\infty}_{(\bar{z}_1,\bar{z}_2)}\circ\pi_{(\bar{z}_1,\bar{z}_2)}$ transforms into $\mathcal{I}^*(\bar{\mathbb{S}}^{\infty}_{(\bar{z}_1,\bar{z}_2)}\circ\pi_{(\bar{z}_1,\bar{z}_2)})$. After some tedious, yet straightforward algebraic manipulations, we have found this pullback is given by
\begin{align}
\mathcal{I}^*(\bar{\mathbb{S}}^{\infty}_{(\bar{z}_1,\bar{z}_2)}\circ\pi_{(\bar{z}_1,\bar{z}_2)})=\mathbb{S}^\infty_{(z_1,z_2)}\circ\pi_{(z_1,z_2)},
\end{align}
where
\begin{align}\label{s_infinity}
\mathbb{S}^\infty_{(z_1,z_2)}(\mathbf{c})=\sum_{k=N_1}^{N_2-1}L_\infty(\mathbf{c}_k,\mathbf{c}_{k+1},\tau k),
\end{align}
and
\begin{align}\label{l_infinity}
L_\infty(\mathbf{c}_k,\mathbf{c}_{k+1},t)=&\bar{L}_\infty(F_{t_1,t}(\mathbf{c}_k),F_{t_1,t}(\mathbf{c}_{k+1}),t)\nonumber\\
=&f(\Phi^t_{t+\tau/2,t}(\mathbf{c}_k),\Phi^t_{t+\tau/2,t+\tau}(F_{t,t+\tau}(\mathbf{c}_{k+1})))\nonumber\\
&+\mathcal{L}^t(\mathbf{c}_k)-\mathcal{L}^{t+\tau}(\mathbf{c}_{k+1})+\left(\int\limits_{t}^{t+\tau}F^*_{s,t+\tau}\mathcal{L}_s\,ds\right)(\mathbf{c}_{k+1})\nonumber\\
&+\left(\epsilon\int\limits_{t+\tau/2}^{t+\tau} \Phi^{t\,*}_{s,t+\tau}l^t_s\,ds\right)(F_{t,t+\tau}(\mathbf{c}_{k+1}))\nonumber\\
&+\left(\epsilon\int\limits_{t}^{t+\tau/2}\Phi^{t\,*}_{s,t} l^t_s\,ds\right)(\mathbf{c}_k).
\end{align}
The notation introduced in this expression is defined as follows. The mapping $\Phi^u_{t,s}$ is the time-dependent flow map of the Hamiltonian vector field with time-dependent Hamiltonian $\epsilon F^*_{t,u}h_t$. In particular,
\begin{align}
\Phi^u_{t,s}=F_{u,t_1}\circ G_{t,s}\circ F_{t_1,u}.
\end{align}
The function $l^u_s$ is given by
\begin{align}
l^u_s=\vartheta(X_{F^*_{s,u}h_s})-F^*_{s,u}h_s.
\end{align}
Finally,
\begin{align}
\mathcal{L}^k=\int_{t_1}^{\tau k}F^*_{s,\tau k}\mathcal{L}_s\,ds.
\end{align}
The functional $\mathbb{S}^\infty_{(z_1,z_2)}$ is therefore a valid exact discrete action, i.e. it satisfies properties (D1)--(D3). Note that the terms $\mathcal{L}^k(\mathbf{c}_k)-\mathcal{L}^{k+1}(\mathbf{c}_{k+1})$ in $L_\infty$ are gauge contributions in the sense that, when summed over $k$, they only contribute a constant to $\mathbb{S}^\infty_{(z_1,z_2)}$. Therefore they can be omitted from $L_\infty$ without affecting the location of critical points. 

\section{\label{sec:truncations_of_l_infinity}Truncations of $\mathbb{S}^\infty_{(z_1,z_2)}$}
While the discrete Euler-Lagrange equations associated with $\mathbb{S}^\infty_{(z_1,z_2)}$ are rigorously satisfied by the integral curves of $X_{\mathcal{H}_t}$ by property (D1), they generally do not serve as a particularly useful numerical integration algorithm. This is because calculating $L_\infty$ is generally very difficult; the flow maps $F_{t,s}$ and $\Phi^u_{t,s}$ must be known in advance. Thus, it is important to have manageable approximations for $\mathbb{S}^\infty_{(z_1,z_2)}$ on hand when developing variational integrators. To this end, we will now present general expressions for the first few terms in $\mathbb{S}^\infty_{(z_1,z_2)}$'s power series in $\epsilon$. The utility of such expressions follows from property (D3): if $\mathbb{S}^\infty_{(z_1,z_2)}$ is replaced by its $l$'th order Maclaurin polynomial in $\epsilon$, the resulting discrete Euler-Lagrange equations will function as a two-step numerical integration algorithm with local $O(\epsilon^{l+1})$ accuracy. We will adopt the convention that $\mathbb{S}^l_{(z_1,z_2)}$ denotes $\mathbb{S}^\infty_{(z_1,z_2)}$'s $l$'th order Maclaurin polynomial in $\epsilon$. 
\\ \\
\noindent {\bf $l=0$:}
\\ \\
When $l=0$, the truncated discrete action is given by
\begin{align}\label{s_0}
\mathbb{S}^0_{(z_1,z_2)}(\mathbf{c})=\sum_{k=N_1}^{N_2-1}L_0(\mathbf{c}_k,\mathbf{c}_{k+1},\tau k),
\end{align}
where
\begin{align}\label{l_0}
L_0(\mathbf{c}_k,\mathbf{c}_{k+1},t)=&f(\mathbf{c}_k,F_{t,t+\tau}(\mathbf{c}_{k+1}))\nonumber\\
&+\mathcal{L}^t(\mathbf{c}_k)-\mathcal{L}^{t+\tau}(\mathbf{c}_{k+1})+\left(\int\limits_{t}^{t+\tau}F^*_{s,t+\tau}\mathcal{L}_s\,ds\right)(\mathbf{c}_{k+1}).
\end{align}
Recall that the function $f$ is defined in Eq.\,(\ref{f}). Also recall that $\mathcal{L}^{\tau k}(\mathbf{c}_k)-\mathcal{L}^{\tau(k+1)}(\mathbf{c}_{k+1})$ is a gauge term and can therefore be omitted. The discrete Euler-Lagrange equations associated with $L_0$ provide a two-step integration algorithm for $X_{\mathcal{H}_t}$ with local $O(\epsilon)$ accuracy. In particular, $\mathbb{S}^0_{(z_1,z_2)}$ serves as an exact discrete action for the unperturbed system $X_{H_t}$. Therefore, $L_0$ can be used to derive variational integrators for \emph{generic} non-canonical Hamiltonian systems without a perturbative structure as follows. Choose $\tau$ to be a small parameter. Then $L_0$ can meaningfully be expanded in a Maclaurin series in $\tau$. From the general theory developed in Ref.\,\cite{Marsden_2001}, it follows that if $L_0$ is replaced with its $n$'th order Maclaurin series in $\tau$, the associated discrete Euler-Lagrange equations will serve as an $(n+1)$'th order integrator.
\\ \\
\noindent {\bf $l=1$:}
\\ \\
In order to derive an expression for $\mathbb{S}^1_{(z_1,z_2)}$, it is necessary to make use of the following identity. Let $K^u_s=F^*_{s,u}h_s$. If $g:M\rightarrow M$ is an arbitrary smooth function, then
\begin{align}\label{expansion_identity}
\Phi^{u\,*}_{t,s}g=&g+\epsilon\int\limits_s^t \Phi^{u\,*}_{a,s}\left(L_{X_{K^u_a}}g\right)\,da\nonumber\\
                         =&g+\epsilon\int\limits_s^t L_{X_{K^u_a}}g\,da+\epsilon^2\int\limits_s^t\int\limits_s^aL_{X_{K^u_b}}L_{X_{K^u_a}}g\,db\,da+O(\epsilon^3),
\end{align}
where $L_X$ denotes the Lie derivative along the vector field $X$. Provided $|t-s|=O(1)$, this identity can be used to obtain an asymptotic expansion of the quantity $\Phi^{u\,*}_{t,s}g$ in powers of $\epsilon$.

Applying this identity to Eq.\,(\ref{l_infinity}), we obtain
\begin{align}\label{s_1}
\mathbb{S}^1_{(z_1,z_2)}(\mathbf{c})=\sum_{k=N_1}^{N_2-1}L_1(\mathbf{c}_k,\mathbf{c}_{k+1},\tau k),
\end{align}
where
\begin{align}\label{l_1}
L_1(\mathbf{c}_k,\mathbf{c}_{k+1},t)=&L_0(\mathbf{c}_k,\mathbf{c}_{k+1},t)\nonumber\\ 
                                                       &+\epsilon\int\limits_{t+\tau/2}^{t+\tau}l^t_s(F_{t,t+\tau}(\mathbf{c}_{k+1}))\,ds+\epsilon\int\limits_{t}^{t+\tau/2}l^t_s(\mathbf{c}_{k})\,ds\nonumber\\
&-\left(\epsilon\int\limits_{t+\tau/2}^{t+\tau}L^{(2)}_{X_{K^t_s}}f\,ds\right)(\mathbf{c}_k,F_{t,t+\tau}(\mathbf{c}_{k+1}))\nonumber\\
&+\left(\epsilon\int\limits_{t}^{t+\tau/2}L^{(1)}_{X_{K^t_s}}f\,ds\right)(\mathbf{c}_k,F_{t,t+\tau}(\mathbf{c}_{k+1})).
\end{align}
The discrete Euler-Lagrange equations associated with $L_1$ furnish a variational integrator for $X_{\mathcal{H}_t}$ with local $O(\epsilon^2)$ accuracy. Keep in mind that the time step $\tau=O(1)$. Practically speaking, for a given value of $\epsilon$, $\tau$ should be significantly less than the perturbation's characteristic bounce time. If $\tau$ \emph{is} chosen to be a small parameter, then these expressions can be expanded in powers of $\tau$. If this expansion were to be performed and $L_1$ were replaced with its $n$'th order Maclaurin polynomial in $\tau$, the resulting discrete Euler-Lagrange equations would yield an integration algorithm with local $O(\tau^{n+1}\epsilon^2)$ accuracy. 
\\ \\
\noindent {\bf $l=2$:}
\\ \\
Upon further application of Eq.\,(\ref{expansion_identity}), the $l=2$ result is given by
\begin{align}\label{s_2}
\mathbb{S}^2_{(z_1,z_2)}(\mathbf{c})=\sum_{k=N_1}^{N_2-1}L_2(\mathbf{c}_k,\mathbf{c}_{k+1},\tau k),
\end{align}
where
\begin{align}\label{l_2}
L_2(\mathbf{c}_k,\mathbf{c}_{k+1},t)=&L_1(\mathbf{c}_k,\mathbf{c}_{k+1},t)\nonumber\\
&-\epsilon^2\int\limits_{t+\tau/2}^{t+\tau}\int\limits_{s}^{t+\tau}L_{X_{K^t_a}}l^t_s(F_{t,t+\tau}(\mathbf{c}_{k+1}))\,da\,ds\nonumber\\
&+\epsilon^2\int\limits_{t}^{t+\tau/2}\int\limits_{t}^{s}L_{X_{K^t_a}}l^t_s(\mathbf{c}_{k})\,da\,ds\nonumber\\
&+\left(\epsilon^2\int\limits_{t+\tau/2}^{t+\tau}\int\limits_{s}^{t+\tau}L^{(2)}_{X_{K^t_a}}L^{(2)}_{X_{K^t_s}}f\,da\,ds\right)(\mathbf{c}_k,F_{t,t+\tau}(\mathbf{c}_{k+1}))\nonumber\\
&+\left(\epsilon^2\int\limits_{t}^{t+\tau/2}\int\limits_{t}^{s}L^{(1)}_{X_{K^t_a}}L^{(1)}_{X_{K^t_s}}f\,da\,ds\right)(\mathbf{c}_k,F_{t,t+\tau}(\mathbf{c}_{k+1}))\nonumber\\
&-\left(\epsilon^2\int\limits_{t}^{t+\tau/2}\int\limits_{t+\tau/2}^{t+\tau}L^{(1)}_{X_{K^t_s}}L^{(2)}_{X_{K^t_a}}f\,da\,ds\right)(\mathbf{c}_k,F_{t,t+\tau}(\mathbf{c}_{k+1})).
\end{align}
The discrete Euler-Lagrange equations associated with $L_2$ furnish a variational integrator for $X_{\mathcal{H}_t}$ with local $O(\epsilon^3)$ accuracy.
\section{\label{sec:symplecticity} Symplecticity}
Fix a non-negative integer $l$. In this section we will discuss the sense in which the numerical integration algorithm associated with $\mathbb{S}^l_{(z_1,z_2)}$ is symplectic. This topic is more subtle than it may first appear because the discrete Euler-Lagrange equations associated with $L_l$ give a two-step integrator for a first-order dynamical system on $M$. 

First we will illustrate the sense in which the integration algorithm associated with $\mathbb{S}^l_{(z_1,z_2)}$ is symplectic on $M\times M$ by applying the methods of Ref.\,\cite{Marsden_2001} in a straightforward manner. Recall that the discrete Euler-Lagrange equations associated with $L_l$ are given by
\begin{align}
\mathbf{d}^{(2)}L_l(\mathbf{c}_{k-1},\mathbf{c}_k,\tau(k-1))+\mathbf{d}^{(1)}L_l(\mathbf{c}_k,\mathbf{c}_{k+1},\tau k)=0.
\end{align}
Thus, we can define a mapping $F^l_k:M\times M\rightarrow M\times M$, where 
\begin{align}
F^l_k(z_1,z_2)=(z_2,f^l_k(z_1,z_2)),
\end{align}
and $f^l_k:M\times M\rightarrow M$ is defined implicitly by the equation
\begin{align}
\mathbf{d}^{(2)}L_l(z_1,z_2,\tau(k-1))+\mathbf{d}^{(1)}L_l(z_2,f^l_k(z_1,z_2),\tau k)=0.
\end{align}
$F^l_k$ allows us to parameterize the space of solutions of the discrete Euler-Lagrange equations by $M\times M$. In particular, it can be used to define a mapping $\mathfrak{C}:M\times M\rightarrow \mathbb{P}$, where $\mathbb{P}$ is the set of mappings $[\tau N_1,\tau N_2]\cap (\tau\mathbb{Z})\rightarrow M$ and
\begin{align}
\mathfrak{C}(z_1,z_2)(\tau N_1)&=z_1\\
\mathfrak{C}(z_1,z_2)(\tau k)&=\pi_1\circ F^l_{k}\circ F^l_{k-1}\circ\ldots\circ F^l_{N_1+1}(z_1,z_2),~~N_1<k\leq N_2
\end{align}
Using this parameterization, we can define the restricted action on $M\times M$, $\hat{\mathbb{S}}^l_{M\times M}:M\times M\rightarrow\mathbb{R}$, where
\begin{align}
\hat{\mathbb{S}}^l_{M\times M}(z_1,z_2)=\sum_{k=N_1}^{N_2-1}L_l(\mathfrak{C}(z_1,z_2)(\tau k),\mathfrak{C}(z_1,z_2)(\tau (k+1)),\tau k).
\end{align} 
The exterior derivative of $\hat{\mathbb{S}}^l_{M\times M}$ is readily found to be
\begin{align}
\mathbf{d}\hat{\mathbb{S}}^l_{M\times M}=\theta_1^{N_1}-F_{N_2-1,N_1}^{l*}\theta_2^{N_2-1},
\end{align}
where $F^l_{N_2-1,N_1}=F^l_{N_2-1}\circ\ldots\circ F^l_{N_1+1}$ and the one-forms $\theta^k_1,\theta^k_2$ are given by
\begin{align}
\theta^k_1(z_1,z_2)(u_{(z_1,z_2)})&=\mathbf{d}L_l(z_1,z_2,\tau k)(T\pi_1(u_{(z_1,z_2)}))\\
\theta^k_2(z_1,z_2)(u_{(z_1,z_2)})&=-\mathbf{d}L_l(z_1,z_2,\tau k)(T\pi_2(u_{(z_1,z_2)})).
\end{align}
Likewise, the second exterior derivative of $\hat{\mathbb{S}}^l_{M\times M}$ is found to be
\begin{align}
\mathbf{d}\mathbf{d}\hat{\mathbb{S}}^l_{M\times M}=0=\mathbf{d}\theta_1^{N_1}-F_{N_2-1,N_1}^{l*}\mathbf{d}\theta_2^{N_2-1}.
\end{align}
It is clear from their definitions that the one-forms $\theta^k_1$ and $\theta^k_2$ differ by an exact differential,
\begin{align}
\theta^k_1(z_1,z_2)-\theta^k_2(z_1,z_2)=\mathbf{d}L_l(z_1,z_2,\tau k).
\end{align}
Therefore we have the conservation law
\begin{align}
F^{l*}_{k,N_1}\omega^k~~\text{is independent of $k$},
\end{align}
where $\omega^k=\mathbf{d}\theta_1^k=\mathbf{d}\theta_2^k$. We have thus shown that the time-dependent flow map $F^l_{k,N_1}$ on $M\times M$ associated with $\mathbb{S}^l_{(z_1,z_2)}$ preserves a time-dependent symplectic form on $M\times M$.

Next we will demonstrate a more refined result on the symplecticity of the integration algorithm associated with $\mathbb{S}^l_{(z_1,z_2)}$. Assume that there exists a time-dependent vector field $X_t$ on $M$ whose associated flow map $\mathcal{F}_{t,s}$ satisfies the discrete Euler-Lagrange equations \emph{exactly}. In other words, $\mathcal{F}_{t,s}$ satisfies the equation
\begin{align}
\mathbf{d}^{(2)}L_l(\mathcal{F}_{\tau(k-1),\tau N_1}(z),\mathcal{F}_{\tau k,\tau N_1}(z),\tau(k-1))\nonumber\\
+\mathbf{d}^{(1)}L_l(\mathcal{F}_{\tau k,\tau N_1}(z),\mathcal{F}_{\tau(k+1),\tau N_1}(z),\tau k)=0.
\end{align}
for each $N_1\leq k\leq N_2$ and $z\in M$. Using $\mathcal{F}_{t,s}$, we can define a mapping $\mathcal{C}:M\rightarrow \mathbb{P}$, where
\begin{align}
\mathcal{C}(z)(\tau k)=\mathcal{F}_{\tau k,\tau N_1}(z).
\end{align}
The mapping $\mathcal{C}$ then naturally leads to the introduction of the restricted action on $M$, $\hat{\mathbb{S}}^l_{M}:M\rightarrow\mathbb{R}$, where
\begin{align}
\hat{\mathbb{S}}^l_{M}(z)=\sum_{k=N_1}^{N_2-1}L_l(\mathcal{F}_{\tau k,\tau N_1}(z),\mathcal{F}_{\tau(k+1),\tau N_1}(z),\tau k).
\end{align}
The identity $\mathbf{d}\mathbf{d}\hat{\mathbb{S}}^l_{M}=0$ then gives the conservation law
\begin{align}
\mathcal{F}_{\tau k,\tau N_1}^*\Omega^k~\text{is independent of $k$},
\end{align}
where $\Omega^k=\mathbf{d}\tilde{\theta}^k_1=\mathbf{d}\tilde{\theta}^k_2$ and
\begin{align}
\tilde{\theta}^k_1(z)&=\mathbf{d}^{(1)}L_l(z,\mathcal{F}_{\tau(k+1),\tau k}(z),\tau k)\\
\tilde{\theta}^k_2(z)&=-\mathbf{d}^{(2)}L_l(\mathcal{F}_{\tau(k-1),\tau k}(z),z,\tau(k-1)).
\end{align}
Note that $\tilde{\theta}^k_1=\tilde{\theta}^k_2$. Thus, we see that the smooth modified system, which is specified by $X_t$, preserves a time-dependent symplectic form on $M$. The sense in which this conservation law applies to the two-step integration algorithm specified by $\mathbb{S}^l_{(z_1,z_2)}$ is as follows. The two step algorithm specified by the discrete Euler-Lagrange equations requires a pair of initial conditions in order to produce a discrete-time trajectory. If the second initial condition is supplied by flowing along $X_t$ for $\tau$ seconds starting from the first initial condition, then the discrete trajectory produced by solving the discrete Euler-Lagrange equations will automatically lie along an integral curve of $X_t$. Thus, provided the second initial condition for our two-step method is chosen carefully, the two-step method is equivalent to the one-step method given by the flow map associated with $X_t$, which we have just shown preserves a time-dependent symplectic form on $M$. This result is consistent with the discussions found in Refs.\,\cite{Ellison_2013, Hairer_1999, Hairer_2006} that explain the advantages of choosing the second initial condition for a two-step integrator using the smooth modified system.

The question of whether or not the vector field $X_t$ exists seems to be incompletely resolved. Using the methods of Hairer, who calls $X_t$ the smooth modified system \cite{Hairer_1999, Hairer_2006}, an asymptotic series for $X_t$ in powers of $\epsilon$ can be developed in a straightforward manner if one assumes the ansatz
\begin{align}\label{ansatz}
X_t=X_{H_t}+\epsilon Y^1_t+\epsilon^2 Y^2_t+\ldots
\end{align}
While, in general, this series diverges, such divergence does not necessarily imply that an $X_t$ with the desired properties fails to exist. It is possible, for instance, that the series\,(\ref{ansatz}) can be resummed in the sense of Borel \cite{Boyd_1999} to give $X_t$. Note that this existence question is not resolved by the result proved in Ref.\,\cite{Kirchgraber_1986}. Indeed, two step variational integrators tend to not be absolutely stable and therefore cannot be treated with the methods of Ref. \cite{Kirchgraber_1986}. 

Regardless of the answer to the existence question, the asymptotic series in Eq.\,(\ref{ansatz}) can often be computed and then truncated at some order. While this truncated vector field will not have a flow map that exactly satisfies the discrete Euler-Lagrange equations, by truncating at a sufficiently high order, it can be made to satisfy the discrete Euler-Lagrange equations with any desired level of accuracy. This fact has already been exploited to improve the stability properties of multi-step variational integration methods in \cite{Ellison_2013}. It would be interesting to also exploit the same fact to develop energy and symplecticity bounds for the discretizations of Hamilton's principle in phase space developed here. We leave this to future consideration. Note that existing proofs of bounded energy errors for variational integrators only apply to discretizations of Hamilton's principle, and not to discretizations of Hamilton's principle in phase space; when the second initial condition supplied to a two-step discretization of Hamilton's principle in phase space is not chosen to lie along an integral curve of $X_t$, there are known examples of poor energy behavior \cite{Squire_2012, Vankerschaver_2013}.

\section{\label{sec:examples}Examples}
In this section we will illustrate the use of our discretizations of Hamilton's principle in phase space to develop variational integrators for a pair of perturbed Hamiltonian systems. In order to illustrate that $L_\infty$ gives discrete Euler-Lagrange equations that are exactly satisfied by the appropriate continuous-time trajectories, we will first treat the harmonic oscillator regarded as a perturbed rigid rotor. In this example, $L_\infty$ is simple to calculate and it is a simple matter to verify that true solutions to the harmonic oscillator differential equation satisfy $L_\infty$'s discrete Euler-Lagrange equations. We will then use the first-order (in $\epsilon$) approximation for $L_\infty$, $L_1$, to derive a variational integrator for a non-canonical Hamiltonian system that describes the nearly-integrable flow of magnetic field lines in a nominally axisymmetric geometry. This second example is less trivial than the Harmonic oscillator in the sense that the field line dynamics are non-linear and $L_\infty$ is impossible to calculate; the computation of $L_1$ is already an onerous task to perform without the aid of symbolic manipulation software.
\\ \\
\noindent \emph{Example 1:}
\\ \\
Consider the canonical Hamiltonian system on $\mathbb{R}^2$ specified by the symplectic form $-\mathbf{d}\vartheta$, with $\vartheta=y\,\mathbf{d}x$, and the Hamiltonian function $\mathcal{H}=y^2/2+\epsilon x^2/2$. Clearly, this system describes the dynamics of a harmonic oscillator with frequency $\sqrt{\epsilon}$. Equivalently, we can regard $\mathcal{H}$ as describing a perturbed rigid rotor, where $H=y^2/2$ describes the unperturbed dynamics of the rotor, and $\epsilon h=\epsilon x^2/2$ describes the perturbation. This second interpretation allows us to employ our discretizations of Hamilton's principle in phase space that are adapted to perturbed problems to develop a variational integrator for this problem.

In order to identify this integrator, we will calculate $L_\infty$ explicitly using Eq.\,(\ref{l_infinity}). The ingredients that enter into such a calculation are (i) introducing an affine connection on $\mathbb{R}^2$, (ii) finding an expression for the function $f(z_1,z_2)$, (iii) finding an expression for the unperturbed flow map $F_{t,s}$, (iv) finding an expression for for $\Phi^u_{t,s}$, and (v) evaluating the necessary time integrals that appear in the expression for $L_\infty$. We will now go through each of these steps in turn.

(i) We will use the obvious connection on $\mathbb{R}^2$ associated with the standard inner product $\left<u,v\right>=u_1v_1+u_2v_2$. Relative to this connection and between any pair of points in $z_1,z_2\in\mathbb{R}^2$, there is a unique geodesic segment equal to the convex hull of $\{z_1,z_2\}$. The unique parameterization of this geodesic segment with parameter $\lambda\in[0,1]$ and orientation $z_1\rightarrow z_2$ is given by
\begin{align}
I(z_2,z_1)(\lambda)=(1-\lambda)\,z_1+\lambda\,z_2.
\end{align}   

(ii) The connection chosen in the previous step renders the computation of $f(z_1,z_2)$ analytically tractable. Indeed, we have
\begin{align}
f(z_1,z_2)&=\int\limits_{I(z_2,z_1)}y\,\mathbf{d}x\nonumber\\
                &=\int_0^1(y_1+\lambda\,(y_2-y_1))\,(x_2-x_1)\,d\lambda\nonumber\\
                &=\frac{1}{2}\,(x_2-x_1)\,(y_1+y_2).
\end{align}

(iii-iv) The unperturbed flow map $F_{t,s}$ is given by
\begin{align}\label{rotor_flow}
F_{t,s}(x,y)=(x+(t-s)\,y,y).
\end{align} 
Determining this flow map is a simple matter because the Hamiltonian underlying the unperturbed dynamics is merely $H=y^2/2$. The flow map of $X_{\mathcal{H}}$, $\mathfrak{F}_{t,s}$, is also simple to identify because the dynamics of a simple harmonic oscillator with frequency $\sqrt{\epsilon}$ are very well understood. Indeed, we have
\begin{align}\label{sho_flow}
&\mathfrak{F}_{t,s}(x,y)=\nonumber\\
&(x\cos(\sqrt{\epsilon}(t-s))+\frac{y}{\sqrt{\epsilon}}\sin(\sqrt{\epsilon}(t-s)),y\cos(\sqrt{\epsilon}(t-s))-\sqrt{\epsilon}x\sin(\sqrt{\epsilon}(t-s))).
\end{align}
On the other hand, the flow map $\Phi^u_{t,s}$ is associated with the time-dependent Hamiltonian 
\begin{align}
\epsilon F^*_{t,u}h_t(x,y)=\epsilon (x+(t-u)\,y)^2/2,
\end{align}
which is not the Hamiltonian for any commonly encountered dynamical system. Nevertheless, it is easy to check that $\Phi^u_{t,s}$ can be expressed in terms of $F_{t,s}$ and $\mathfrak{F}_{t,s}$ according to
\begin{align}
\Phi^u_{t,s}=F_{u,t}\circ\mathfrak{F}_{t,s}\circ F_{s,u}.
\end{align}
Thus, between Eqs.\,(\ref{rotor_flow}) and (\ref{sho_flow}) we have identified an explicit expression for $\Phi^u_{t,s}$ (that we will not display).

(v) Finally, we can calculate $L_\infty$ by directly evaluating Eq.\,(\ref{l_infinity}). Modulo gauge terms, $L_\infty$ is given by
\begin{align}
L_\infty(z_1,z_2)=&-\frac{1}{2}(y_2x_1-y_1x_2)\cos(\sqrt{\epsilon}\tau)\nonumber\\
                             &-\frac{1}{2}\left(\frac{1}{\sqrt{\epsilon}}y_2y_1+\sqrt{\epsilon}x_2x_1\right)\sin(\sqrt{\epsilon}
                             \tau).
\end{align}
It is simple to verify the the discrete Euler-Lagrange equations that follow from this expression for $L_\infty$ are exactly satisfied by the solution to the harmonic oscillator differential equation. This is true regardless of how large or small the time step $\tau$ is chosen.

In this case, $L_\infty$ can technically be expanded in powers of $\epsilon$ regardless of the value of $\tau$. This follows from the fact that the radius of convergence of the Maclaurin series of either $\sin(x)$ or $\cos(x)$ is infinite. However, when $\sqrt{\epsilon}\tau\ll 1$ these series converge much more rapidly than when $\sqrt{\epsilon}\tau\geq 1$. Thus, we expect that truncating $L_\infty$'s Maclaurin series in $\epsilon$ after only a few terms will lead to a reasonably-accurate integrator for this perturbed rigid rotor only when $\sqrt{\epsilon}\tau\ll 1$, which is precisely the condition that $\tau$ be much less than the characteristic bounce time $1/\sqrt{\epsilon}$.
\\ \\
\emph{Example 2:}
\\ \\
Next, we will summarize the application of the discretizations developed in this work to a non-trivial non-canonical perturbed Hamiltonian system. This system's phase space is $\mathbb{R}^2$ equipped with the non-canonical symplectic form $-\mathbf{d}\vartheta$, where
\begin{align}
\vartheta=(x^2+y^2)\,(y\,\mathbf{d}x-x\,\mathbf{d}y).
\end{align} 
The Hamiltonian, which is time-dependent and periodic, is given by $\mathcal{H}_t=H+\epsilon h_t$, where
\begin{align}
H&=\frac{2}{9} (x^2+y^2)^3\\
h_t&=x\sin(t)+x^{2}\sin(t).
\end{align}
This Hamiltonian system can be regarded as a model for magnetic field line flow in a nominally axisymmetric geometry with small resonant perturbations \cite{Cary_1983}. In this interpretation, the time variable is identified with the toroidal angle, while $x$ and $y$ are identified with a set of Cartesian coordinates in the poloidal plane centered on the unperturbed magnetic axis. We will demonstrate that the variational integrators for this system given by $L_0$ and $L_1$ have local $O(\epsilon)$ and $O(\epsilon^2)$ accuracy, respectively. Then we will make some qualitative remarks on the ability of the $L_1$ integrator to resolve second- and higher-order islands. The time step for the $L_0$ and $L_1$ integrators will be set equal to $2\pi$. Thus, these integrators function as the zero'th and first order $t=0$ Poincar\'e maps.

In order to find an expression for $L_1$, a connection must be chosen; $f(z_1,z_2)$ must be calculated; and the unperturbed flow map must be found. With knowledge of these quantities, Eq.(\ref{l_1}) can be evaluated directly to find the desired expression. As in the previous example, we adopt the natural flat connection on $\mathbb{R}^2$. Relative to this connection, the function $f(z_1,z_2)$ is given by
\begin{align}
f(z_1,z_2)=\frac{1}{3}(x_2y_1-x_1y_2)(x_1^2+x_1x_2+x_2^2+y_1^2+y_1y_2+y_2^2).
\end{align} 
Finally, the flow map of the dynamical system defined by the Hamiltonian function $H$ is given by
\begin{align}
F_{t,s}(x,y)=\bigg(&x\cos\left(\frac{1}{3}(x^2+y^2)(t-s)\right)+y\sin\left(\frac{1}{3}(x^2+y^2)(t-s)\right),\nonumber\\
                     -&x\sin\left(\frac{1}{3}(x^2+y^2)(t-s)\right)+y\cos\left(\frac{1}{3}(x^2+y^2)(t-s)\right)\bigg).
\end{align}
We will not display the result of using these quantities to calculate $L_1$ because the resulting expression has many terms, but we must emphasize that this tedious calculation is readily performed using a symbolic manipulation tool such as \textit{Mathematica}. The laborious task of differentiating such a complicated discrete Lagrangian can be handled by either computing the needed derivatives symbolically before running a simulation or by employing an automatic differentiation tool such as ADOLC\,\cite{Griewank_2000,Walther_2012} at runtime. By employing automatic differentiation tools, implementation and testing of lengthy expressions appearing in the discrete euler-lagrange equations and nonlinear solve can be avoided. On the other hand, precomputing derivatives may lead to shorter run times. 

\begin{figure}
\includegraphics{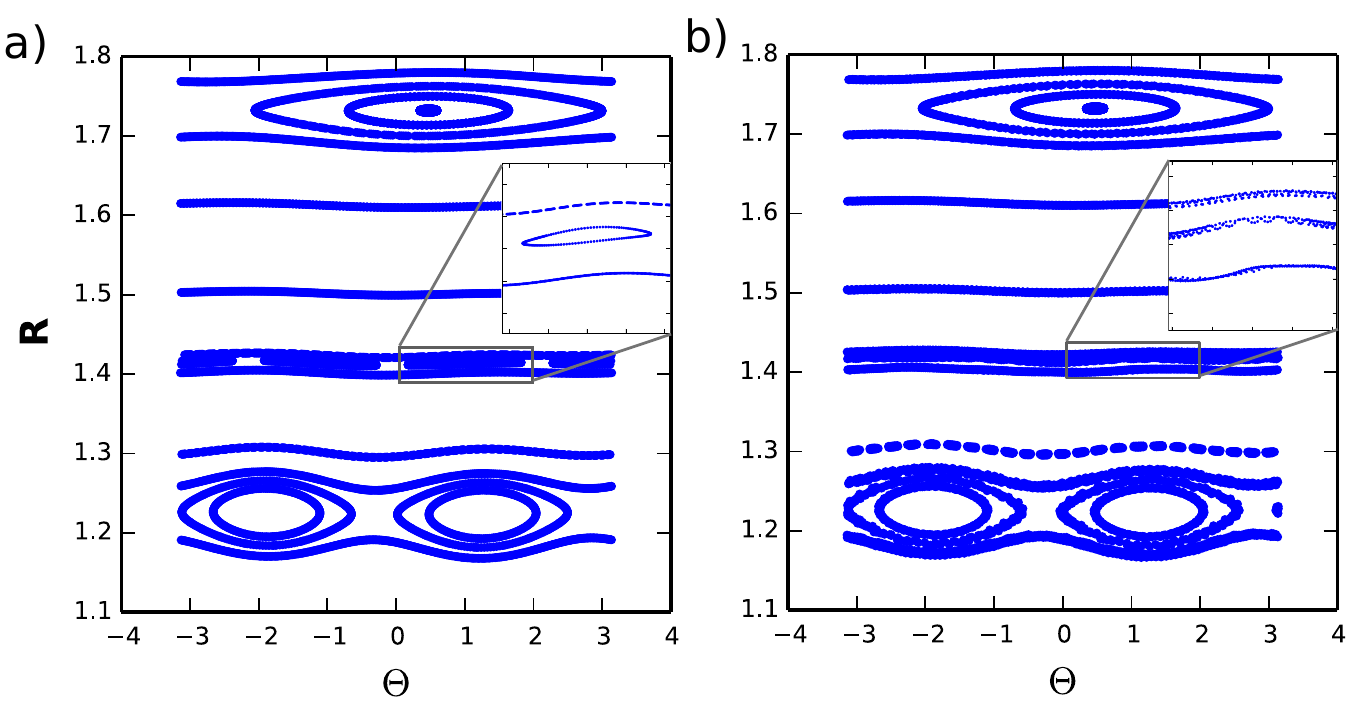}
\caption{\label{fig:two_rmps_poincare}(a) The $t=0$ Poincar\'e section for $\mathcal{H}_t$ calculated using a Runge-Kutta integrator with very fine temporal resolution. (b) The same Poincar\'e section computed using the discrete Euler-Lagrange equations associated with $L_1$. In each case, $\epsilon=.0075$ and the variables $R$ and $\Theta$ denote the standard polar coordinates on $\mathbb{R}^2$. The large first-order islands are reproduced well by the $L_1$-integrator, while, as the highlighted portions of the figures indicate, higher-order islands are not properly reproduced \cite{Hunter_2007}. } 
\end{figure}

Figure\,\ref{fig:two_rmps_poincare} shows the results of using the discrete Euler-Lagrange equations associated with $L_1$ to generate the $t=0$ Poincar\'e section for the dynamical system specified by $\mathcal{H}_t$. Notably, the $L_1$-integrator reproduced the first-order islands very well. This can be checked upon noting that the unperturbed frequency as a function of the distance $R$ from the origin in $\mathbb{R}^2$ is given by $\omega(R)=R^2/3$; first-order perturbation theory predicts an island chain at $R_1=\sqrt{3}$ and another at $R_2=\sqrt{3/2}$. On the other hand, as the highlighted portion of the figure indicates, higher-order island chains were not captured correctly by $L_1$. This shortcoming is to be expected in light of the fact that the flow map associated with $\mathcal{H}_t$ only satisfies the discrete Euler-Lagrange equations associated with $L_1$ up to terms second-order in $\epsilon$. The ``incoherent" fine-scale structure present in the $L_1$ integration came as a result of the onset of parasitic modes that generally plague multistep integration methods \cite{Hairer_1999}. The same parasitic modes completely destabilized the $L_1$ integration run shown in Figure\,\ref{fig:two_rmps_poincare} after several tens of thousands of iterations. Smaller values of epsilon can perform larger numbers of iterations before being overtaken by parasitic modes.

\begin{figure}
\includegraphics{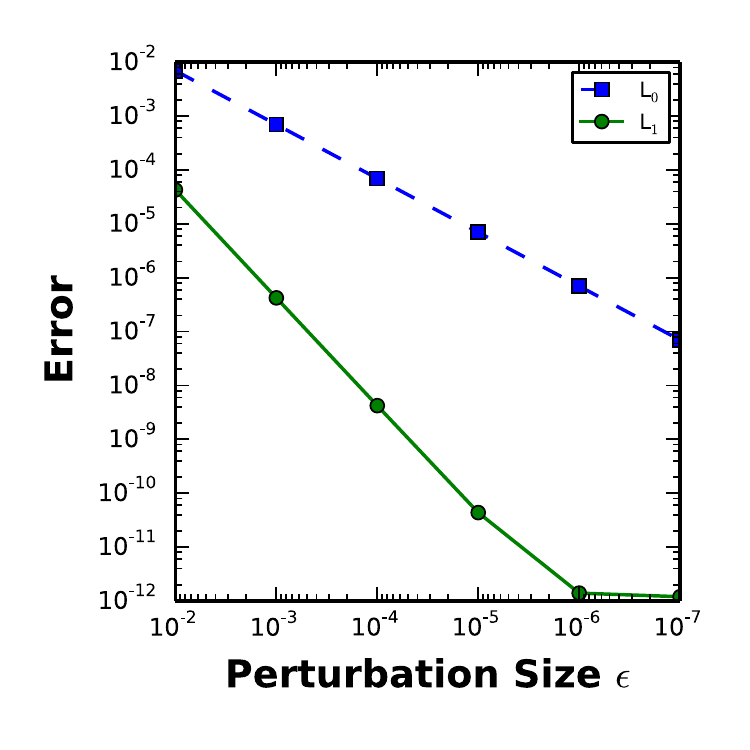}
\caption{\label{fig:two_rmps_convergence} ``Error" denotes the distance between the predictions of a well-converged $2\pi$-second Runge-Kutta integration and a single iteration of the $L_0$ or $L_1$ integrator. Each integration method was initialized with the same initial condition.}
\end{figure}

Figure \,\ref{fig:two_rmps_convergence} illustrates the $O(\epsilon)$ and $O(\epsilon^2)$ errors of the integrators provided by $L_0$ and $L_1$, respectively. The $L_1$ integrator decreases in error quadratically with $\epsilon$ until reaching the error tolerance of the nonlinear Newton-Rhapson solver of $10^{-12}$. Decreasing $\epsilon$ beyond $10^{-6}$ therefore shows no further improvement in the error of $L_1$. 

\section{\label{sec:discussion}Discussion}
In the preceding sections, we presented and applied the first discretizations of Hamilton's principle in phase space that are adapted to perturbed noncanonical Hamiltonian systems. Notably, these discretizations function as variational integrators with $O(1)$ time steps and local $O(\epsilon^N)$ accuracy, where $N$ is any desired nonnegative integer. Moreover, for each discretization, our expression for the associated discrete Lagrangian is manifestly coordinate independent. This coordinate independence has been achieved by introducing an arbitrary affine connection on the phase space. Thus, our discretizations may prove to be useful for constructing variational integrators on manifolds.

We have also shown that if these two-step integrators are initialized using the smooth modified system studied in backward error analysis, then they function as one-step algorithms that preserve symplectic forms on the phase space $M$. We expect that this result will play an important role in the backward error analysis of the variational discretizations developed here. For instance, using the fact that the smooth modified system preserves a symplectic form on $M$, it should be possible to show that the method for choosing the second initial condition for two-step methods championed in \cite{Hairer_1999, Ellison_2013} leads to better energy behavior than a more conventional Runge-Kutta-based initialization.

While the integrators identified here formally apply to any perturbed non-canonical Hamiltonian system, they are \emph{much easier} to apply to nearly-integrable systems because the unperturbed flow map can often be determined analytically in these cases. When the unperturbed flow map is analytically unattainable, it would have to be determined numerically. This would entail devising some numerical scheme for evaluating (at least in an approximate sense) the various integrations along unperturbed orbits that appear in the discrete Lagrangians presented in Section \ref{sec:truncations_of_l_infinity}. Whether or not such a scheme exists that does not involve prohibitively large computational overhead is currently unknown to us. We leave investigating this issue to future work.

A theoretical application of the integrators developed here that we will pursue in the future is \emph{coarse-graining} Hamilton's principle in phase space. Specifically, we would like to derive the stochastic action mentioned in Ref.\,\cite{Burby_2013} by directly manipulating Hamilton's principle in phase space. Our hope is that this result will follow by appropriately rescaling time and then looking at the behavior of $\mathbb{S}^2_{(z_1,z_2)}$ (Eq. (\ref{s_2})) as $\epsilon\rightarrow 0$.

\section{Acknowledgements}
The authors would like to express their gratitude to A. I. Zhmoginov for his help in editing this manuscript. This work was supported by the U.S. Department of Energy under contract DE-AC02-09CH11466.





\bibliographystyle{model1-num-names}

\begin{thebibliography}{32}
\expandafter\ifx\csname natexlab\endcsname\relax\def\natexlab#1{#1}\fi
\providecommand{\url}[1]{\texttt{#1}}
\providecommand{\href}[2]{#2}
\providecommand{\path}[1]{#1}
\providecommand{\DOIprefix}{doi:}
\providecommand{\ArXivprefix}{arXiv:}
\providecommand{\URLprefix}{URL: }
\providecommand{\Pubmedprefix}{pmid:}
\providecommand{\doi}[1]{\href{http://dx.doi.org/#1}{\path{#1}}}
\providecommand{\Pubmed}[1]{\href{pmid:#1}{\path{#1}}}
\providecommand{\bibinfo}[2]{#2}
\ifx\xfnm\relax \def\xfnm[#1]{\unskip,\space#1}\fi
\bibitem[{Landau and Lifshitz(1976)}]{Landau_1976}
\bibinfo{author}{L.~D. Landau}, \bibinfo{author}{E.~M. Lifshitz},
  \bibinfo{title}{Mechanics}, \bibinfo{publisher}{Elsevier
  Butterworth-Heinemann}, \bibinfo{year}{1976}.
\bibitem[{Arnold(1989)}]{Arnold_1989}
\bibinfo{author}{V.~I. Arnold}, \bibinfo{title}{Mathematical Methods of
  Classical Mechanics}, \bibinfo{publisher}{Springer}, \bibinfo{year}{1989}.
\bibitem[{Abraham and Marsden(2008)}]{Abraham_2008}
\bibinfo{author}{R.~Abraham}, \bibinfo{author}{J.~Marsden},
  \bibinfo{title}{Foundations of Mechanics}, AMS Chelsea publishing,
  \bibinfo{publisher}{American Mathematical Soc.}, \bibinfo{year}{2008}.
\bibitem[{Holm et~al.(1998)Holm, Marsden, and Ratiu}]{Holm_1998}
\bibinfo{author}{D.~D. Holm}, \bibinfo{author}{J.~E. Marsden},
  \bibinfo{author}{T.~S. Ratiu},
\newblock \bibinfo{title}{The {E}uler-{P}oincar\'e equations and semidirect
  products with applications to continuum theories},
\newblock \bibinfo{journal}{Adv. Math} \bibinfo{volume}{137}
  (\bibinfo{year}{1998}) \bibinfo{pages}{1}.
\bibitem[{Brizard and Hahm(2007)}]{Brizard_2007}
\bibinfo{author}{A.~J. Brizard}, \bibinfo{author}{T.~S. Hahm},
\newblock \bibinfo{title}{Foundations of nonlinear gyrokinetic theory},
\newblock \bibinfo{journal}{Rev. Mod. Phys.} \bibinfo{volume}{79}
  (\bibinfo{year}{2007}) \bibinfo{pages}{421--468}.
\bibitem[{Cary and Brizard(2009)}]{Cary_2009}
\bibinfo{author}{J.~Cary}, \bibinfo{author}{A.~Brizard},
\newblock \bibinfo{title}{Hamiltonian theory of guiding-center motion},
\newblock \bibinfo{journal}{Rev. Mod. Phys.} \bibinfo{volume}{81}
  (\bibinfo{year}{2009}) \bibinfo{pages}{693}.
\bibitem[{Marsden and West(2001)}]{Marsden_2001}
\bibinfo{author}{J.~E. Marsden}, \bibinfo{author}{M.~West},
\newblock \bibinfo{title}{Discrete mechanics and variational integrators},
\newblock \bibinfo{journal}{Acta Numer.} \bibinfo{volume}{10}
  (\bibinfo{year}{2001}) \bibinfo{pages}{357}.
\bibitem[{Aref(2007)}]{Aref_2007}
\bibinfo{author}{H.~Aref},
\newblock \bibinfo{title}{Point vortex dynamics: A classical mathematics
  playground},
\newblock \bibinfo{journal}{J. Math. Phys} \bibinfo{volume}{48}
  (\bibinfo{year}{2007}) \bibinfo{pages}{5401}.
\bibitem[{Qin and Guan(2008)}]{Qin_2008}
\bibinfo{author}{H.~Qin}, \bibinfo{author}{X.~Guan},
\newblock \bibinfo{title}{Variational symplectic integrator for long-time
  simulations of the guiding-center motion of charged particles in general
  magnetic fields},
\newblock \bibinfo{journal}{Phys. Rev. Lett.} \bibinfo{volume}{100}
  (\bibinfo{year}{2008}) \bibinfo{pages}{035006}.
\bibitem[{Qin et~al.(2009)Qin, Guan, and Tang}]{Qin_2009}
\bibinfo{author}{H.~Qin}, \bibinfo{author}{X.~Guan}, \bibinfo{author}{W.~M.
  Tang},
\newblock \bibinfo{title}{Variational symplectic algorithm for guiding center
  dynamics and its application in tokamak geometry},
\newblock \bibinfo{journal}{Phys. Plasmas} \bibinfo{volume}{16}
  (\bibinfo{year}{2009}) \bibinfo{pages}{042510}.
\bibitem[{Rowley and Marsden(2002)}]{Rowley_2002}
\bibinfo{author}{C.~W. Rowley}, \bibinfo{author}{J.~E. Marsden},
\newblock \bibinfo{title}{Variational integrators for degenerate {L}agrangians,
  with application to point vortices},
\newblock \bibinfo{journal}{41st IEEE Conference on Decision and Control}
  \bibinfo{volume}{40} (\bibinfo{year}{2002}) \bibinfo{pages}{1521}.
\bibitem[{Li et~al.(2011)Li, Qin, Pu, Xie, and Fu}]{Li_2011}
\bibinfo{author}{J.~Li}, \bibinfo{author}{H.~Qin}, \bibinfo{author}{Z.~Pu},
  \bibinfo{author}{L.~Xie}, \bibinfo{author}{S.~Fu},
\newblock \bibinfo{title}{Variational symplectic algorithm for guiding center
  dynamics in the inner magnetosphere},
\newblock \bibinfo{journal}{Phys. Plasmas} \bibinfo{volume}{18}
  (\bibinfo{year}{2011}) \bibinfo{pages}{052902}.
\bibitem[{Lall and West(2006)}]{Lall_2006}
\bibinfo{author}{S.~Lall}, \bibinfo{author}{M.~West},
\newblock \bibinfo{title}{Discrete variational {H}amiltonian mechanics},
\newblock \bibinfo{journal}{J. Phys. A: Math. Gen.} \bibinfo{volume}{39}
  (\bibinfo{year}{2006}) \bibinfo{pages}{5509}.
\bibitem[{Leok and Zhang(2011)}]{Leok_2011}
\bibinfo{author}{M.~Leok}, \bibinfo{author}{J.~Zhang},
\newblock \bibinfo{title}{Discrete {H}amiltonian variational integrators},
\newblock \bibinfo{journal}{IMA J. Numer. Anal.} \bibinfo{volume}{31}
  (\bibinfo{year}{2011}) \bibinfo{pages}{1497}.
\bibitem[{Farr(2009)}]{Farr_2009}
\bibinfo{author}{W.~M. Farr},
\newblock \bibinfo{title}{Variational integrators for almost integrable
  systems},
\newblock \bibinfo{journal}{Celest. Mech. Dyn. Astron.} \bibinfo{volume}{103}
  (\bibinfo{year}{2009}) \bibinfo{pages}{105}.
\bibitem[{McLachlan(1995)}]{Mclachlan_1995}
\bibinfo{author}{R.~I. McLachlan},
\newblock \bibinfo{title}{Composition methods in the presence of small
  parameters},
\newblock \bibinfo{journal}{BIT Numer. Math.} \bibinfo{volume}{35}
  (\bibinfo{year}{1995}) \bibinfo{pages}{258}.
\bibitem[{Chambers and Murison(2000)}]{Chambers_2000}
\bibinfo{author}{J.~E. Chambers}, \bibinfo{author}{M.~A. Murison},
\newblock \bibinfo{title}{Pseudo-high-order symplectic integrators},
\newblock \bibinfo{journal}{Astron. J.} \bibinfo{volume}{119}
  (\bibinfo{year}{2000}) \bibinfo{pages}{425}.
\bibitem[{Laskar and Robutel(2001)}]{Laskar_2001}
\bibinfo{author}{J.~Laskar}, \bibinfo{author}{P.~Robutel},
\newblock \bibinfo{title}{High order symplectic integrators for perturbed
  {H}amiltonian systems},
\newblock \bibinfo{journal}{Celest. Mech. Dyn. Astron.} \bibinfo{volume}{80}
  (\bibinfo{year}{2001}) \bibinfo{pages}{39}.
\bibitem[{Hairer(1999)}]{Hairer_1999}
\bibinfo{author}{E.~Hairer},
\newblock \bibinfo{title}{Backward error analysis for multistep methods},
\newblock \bibinfo{journal}{Numer. Math} \bibinfo{volume}{84}
  (\bibinfo{year}{1999}) \bibinfo{pages}{199--232}.
\bibitem[{Hairer et~al.(2006)Hairer, Lubich, and Wanner}]{Hairer_2006}
\bibinfo{author}{E.~Hairer}, \bibinfo{author}{C.~Lubich},
  \bibinfo{author}{G.~Wanner}, \bibinfo{title}{Geometric Numerical
  Integration}, \bibinfo{publisher}{Springer}, \bibinfo{year}{2006}.
\bibitem[{Ellison et~al.(2013)Ellison, Burby, Finn, Qin, and
  Tang}]{Ellison_2013}
\bibinfo{author}{C.~L. Ellison}, \bibinfo{author}{J.~W. Burby},
  \bibinfo{author}{J.~M. Finn}, \bibinfo{author}{H.~Qin},
  \bibinfo{author}{W.~M. Tang}, \bibinfo{title}{Initializing and stabilizing
  multistep algorithms for modeling dynamical systems}, \bibinfo{year}{2013}.
  \bibinfo{note}{{J. Comput. Phys.} (to be submitted)}.
\bibitem[{Dahlquist(1956)}]{Dahlquist_1956}
\bibinfo{author}{G.~Dahlquist},
\newblock \bibinfo{title}{Convergence and stability in the numerical
  integration of ordinary differential equations},
\newblock \bibinfo{journal}{Math. Scand.} \bibinfo{volume}{4}
  (\bibinfo{year}{1956}) \bibinfo{pages}{33--53}.
\bibitem[{Postnikov(2001)}]{Postnikov_2001}
\bibinfo{author}{M.~M. Postnikov}, \bibinfo{title}{Geometry VI: Riemannian
  Geometry}, Encyclopaedia of Mathematical Sciences,
  \bibinfo{publisher}{Springer}, \bibinfo{year}{2001}. \URLprefix
  \url{http://books.google.com/books?id=1kyUImXf8U0C}.
\bibitem[{Boyd(1999)}]{Boyd_1999}
\bibinfo{author}{J.~P. Boyd},
\newblock \bibinfo{title}{The devil's invention: Asymptotic, superasymptotic,
  and hyperasymptotic series},
\newblock \bibinfo{journal}{Acta Appl. Math.} \bibinfo{volume}{56}
  (\bibinfo{year}{1999}) \bibinfo{pages}{98}.
\bibitem[{Kirchgraber(1986)}]{Kirchgraber_1986}
\bibinfo{author}{U.~Kirchgraber},
\newblock \bibinfo{title}{Multi-step methods are essentially one-step methods},
\newblock \bibinfo{journal}{Numer. Math} \bibinfo{volume}{48}
  (\bibinfo{year}{1986}) \bibinfo{pages}{85--90}.
\bibitem[{Squire et~al.(2012)Squire, Qin, and Tang}]{Squire_2012}
\bibinfo{author}{J.~Squire}, \bibinfo{author}{H.~Qin}, \bibinfo{author}{W.~M.
  Tang},
\newblock \bibinfo{title}{Gauge properties of the guiding center variational
  symplectic integrator},
\newblock \bibinfo{journal}{Phys. Plasmas} \bibinfo{volume}{19}
  (\bibinfo{year}{2012}) \bibinfo{pages}{052501}.
\bibitem[{Vankerschaver and Leok(2013)}]{Vankerschaver_2013}
\bibinfo{author}{J.~Vankerschaver}, \bibinfo{author}{M.~Leok},
\newblock \bibinfo{title}{A novel formulation of point vortex dynamics on the
  sphere: Geometrical and numerical aspects},
\newblock \bibinfo{journal}{J. Nonlinear Sci.} \bibinfo{volume}{July}
  (\bibinfo{year}{2013}) \bibinfo{pages}{01--37}.
\bibitem[{Cary and Littlejohn(1983)}]{Cary_1983}
\bibinfo{author}{J.~R. Cary}, \bibinfo{author}{R.~G. Littlejohn},
\newblock \bibinfo{title}{Noncanonical {H}amiltonian mechanics and its
  application to magnetic field line flow},
\newblock \bibinfo{journal}{Ann. Phys.} \bibinfo{volume}{151}
  (\bibinfo{year}{1983}) \bibinfo{pages}{1}.
\bibitem[{Griewank and Walther(2009)}]{Griewank_2000}
\bibinfo{author}{A.~Griewank}, \bibinfo{author}{A.~Walther},
  \bibinfo{title}{Evaluating Derivatives: Principles and Techniques},
  \bibinfo{publisher}{SIAM: Frontiers in Mathematics}, \bibinfo{year}{2009}.
\bibitem[{Walther and Griewank(2012)}]{Walther_2012}
\bibinfo{author}{A.~Walther}, \bibinfo{author}{A.~Griewank},
  \bibinfo{title}{Getting started with ADOL-C, from Combinatorial Scientific
  Computing}, \bibinfo{publisher}{Chapman-Hall CRC Computational Science},
  \bibinfo{year}{2012}.
\bibitem[{Hunter(2007)}]{Hunter_2007}
\bibinfo{author}{J.~D. Hunter},
\newblock \bibinfo{title}{Matplotlib: A 2d graphics environment},
\newblock \bibinfo{journal}{Computing In Science \& Engineering}
  \bibinfo{volume}{9} (\bibinfo{year}{2007}) \bibinfo{pages}{90--95}.
\bibitem[{Burby et~al.(2013)Burby, Zhmoginov, and Qin}]{Burby_2013}
\bibinfo{author}{J.~W. Burby}, \bibinfo{author}{A.~I. Zhmoginov},
  \bibinfo{author}{H.~Qin},
\newblock \bibinfo{title}{Hamiltonian mechanics of stochastic acceleration},
\newblock \bibinfo{journal}{Phys. Rev. Lett.} \bibinfo{volume}{111}
  (\bibinfo{year}{2013}) \bibinfo{pages}{195001}.

\end{thebibliography}


\providecommand{\noopsort}[1]{}\providecommand{\singleletter}[1]{#1}%








\end{document}